\def\cm2{cm$^{-2}$}

\def\c2{C~{\sc ii}}
\def\c4{C~{\sc iv}}
\def\fe2{Fe~{\sc ii}}
\def\fe3{Fe~{\sc iii}}
\def\mg1{Mg~{\sc i}}
\def\mg2{Mg~{\sc ii}}
\def\si2{Si~{\sc ii}}
\def\si4{Si~{\sc iv}}
\def\al2{Al~{\sc ii}}
\def\al3{Al~{\sc iii}}
\def\o1{O~{\sc i}}
\def\n1{N~{\sc i}}
\def\h1{H~{\sc i}}

\def\approxlt{\mathrel{\spose{\lower 3pt\hbox{$\sim$}}
        \raise 2.0pt\hbox{$<$}}}
\def\approxgt{\mathrel{\spose{\lower 3pt\hbox{$\sim$}}
        \raise 2.0pt\hbox{$>$}}}

\documentclass[article]{has40}
\usepackage{graphicx}
\usepackage{amssymb}
\tabletypesize{\normalsize}  

\shortauthors{Welch \& Foster}
\shorttitle{OGLE-III RR1 Lightcurve Modulations}

\begin{document}
\large    
\pagenumbering{arabic}
\setcounter{page}{116}

\title{A Study of RR1 Lightcurve Modulation in OGLE-III Bulge Time-series}

\author{{\noindent Douglas L. Welch{$^{\rm 1}$} and Grant Foster{$^{\rm 2}$}\\
\\
{\it (1) Department of Physics and Astronomy, McMaster University, Hamilton, Ontario, Canada\\
(2) Tempo Analytics, Garland, Maine, USA} 
}
}

\email{(1) welch@physics.mcmaster.ca (2) twistor9@gmail.com}

\begin{abstract}
We report the results of our study of lightcurve modulation
in a sample of 493 RR1 variables from the OGLE-III survey of galactic bulge fields.
Each object in this list has 1500 or more I-band observations.
We compare our findings with earlier studies regarding lightcurve modulation in LMC  
and galactic field RR1 stars. We also report the discovery of the modulated-Blazhko RR1
star OGLE-BLG-RRLYR-03825 which has a Blazhko period of 16.469 d which itself is modulated
with a period of 339.2 d.
\end{abstract}

\section{Introduction and Motivation}
The first-overtone radial-pulsating RR Lyrae stars, known as RRc type, or RR1 in
the more physically-intuitive mode-based classification system of \cite{2000ApJ...542..257A}, are
relatively poorly studied compared to their higher-amplitude fundamental-mode
counterparts, the RRab (also known as RR0) variables. In part, this situation is a legacy 
of the many decades of photographic variable star surveys where their smaller
photometric amplitudes and more sinusoidal lightcurves resulted in fewer
discoveries. These same factors also made it difficult to detect and study
changes in lightcurve behavior. 

It is only with the advent of wide-field, long-term photometric surveys that
large numbers of galactic field RR1 stars have been identified without significant
selection biases. Such surveys allow the prospect of answering questions
such as:
\begin{itemize}
\item How does the period/amplitude modulation of RR1 stars compare to that
of RR0 stars?
\item Is there a period dependence of the period/amplitude modulation?
\item Is there any evidence of binarity in and of the power spectra of
RR1 stars and do they share the peculiar lack of detected binarity of
tens of thousands of RR0 stars?
\item Is there evidence of for period-doubling in any RR1 time series?
\item Are any more complicated or unexpected behaviors seen only in the RR0
class?
\end{itemize}

\section{Photometric Time-series Source and Characteristics}

The photometric precision and reliable cadence of Kepler satellite mission data
would seem a natural choice for studying RR1 lightcurve behavior and
the power of such data has been demonstrated by \cite{2012arXiv1208.4251M}.
However, the Kepler field contains only four known RR1 stars and consequently ground-based
surveys are still required to evaluate correlations of lightcurve behavior
with other observable characteristics such as period and metallicity.

The analysis in this paper is based upon the massive photometric survey
of galactic bulge fields by \cite{2011AcA....61....1S} who reported
11,756 RR0 and 4,989 RR1 variables, respectively. Two of their key findings
were that 75\% of RR1 in the 0.33-0.45 d period range had detectable
period changes and that ``Blazhko''-like lightcurve modulation was found in
8\% of the RR1 sample. 

The data sample provided by \cite{2011AcA....61....1S} is especially
suitable for further time-series analysis for several reasons. First,
the signal-to-noise ratio of their I-band lightcurves is typically
50:1. Second, the photometry was collected over 8 years or more, providing
unusually long and uniform quality time-series. Third, galactic bulge
RR1 stars are in heavily-crowded regions and the excellent seeing
at the site and image scale of 0.26 arcsec/pixel are critically-important
and unrivaled by other surveys. The single, unavoidable downside
of studying variables in the galactic bulge is the
annual interruption of observations when the Sun passes through
the region.

There are a handful of more detailed studies of RR0 lightcurve modulation
in the literature. \cite{2007MNRAS.377.1263S} analysed galactic field 
RR1 stars which had V-band lightcurves available from ASAS. Of the
756 RR1 stars studied, they found 49 (i.e. 6.5\%) exhibited the
Blazhko effect. They also noted that 15 of the 756 RR1 stars (i.e. 2\%)
showed period change. Their most intriguing finding was a single RR1
star, LS Her, which appeared to show a double-Blazhko modulation.
The properties of the double modulation of LS Her were more completely
investigated by \cite{2008MNRAS.387..783W} who found that it had a
(quite short) Blazhko period of 12.75 days which was itself modulated
with a period of 109 days. At the time, this was the only known
instance of such lightcurve modulation.

\cite{2013JAVSO..41...75P} investigated period changes of RR1 stars
which had recorded brightness maxima in the GOES database. Using
41 stars for which sufficient historical times of maxima existed,
they found lower rates of period change for RR1 stars with periods
shorted than 0.25 days. They also noted that one of the well-observed
galactic field RR1 stars, SX UMa, had a very unstable period.

Times of maximum are not ideal for studying the lightcurve evolution
of RR1 stars because of both lightcurves features near
maximum and the relatively large fraction of pulsation phase spent
near maximum brightness. When sufficient photometry exists, it is
always better to incorporate the whole lightcurve into the analysis.

\section{Method}
We chose to restrict our investigation to the RR1 time-series
from \cite{2011AcA....61....1S} which had at least 1,500 I-band
measurements available. Our selection criterion resulted in 493
lightcurves.

The photometric time-series were initially analyzed using the sequential CLEANest 
algorithm of \cite{1995AJ....109.1889F}. ``Sequential CLEANest'' is very similar 
to the more common method of sequential pre-whitening in which, at each iteration,
the strongest frequency in the Fourier spectrum is added to the frequency list 
(a set of discrete frequencies with which to model the underlying signal), and its 
effect removed from the data. Then a Fourier spectrum is computed from the residuals. 
In contrast to sequential pre-whitening, the subsequent instances of CLEANest 
allow all frequencies to vary whenever a new one is added to the frequency list, 
optimizing the frequency set at each step. Also, the discrete frequencies constituting 
the frequency set can be restored to the spectrum simply by plotting a line to 
represent them, which is most easily done by plotting an amplitude rather than power 
spectrum, making the line's height equal to the amplitude of that frequency in a 
best-fit model to the data.

In most cases the refinement of frequencies which distinguishes sequential CLEANest 
from sequential pre-whitening brings about only small changes to previously determined 
frequencies. However, when periodic fluctuations are subject to very slow modulation, 
previously determined frequencies can change substantially and CLEANest will often 
settle on a set of very closely-spaced frequencies - often much more closely-spaced 
than the native resolution of the Fourier spectrum. These clusters of frequencies make
for a poor - sometimes very poor - representation of the Fourier spectral density, but a 
superior model of the signal underlying the data.

The representation of the data from the frequencies determined by CLEANest can then 
be used to quantify the time-varying period, amplitude, and phase of the underlying signal, 
i.e. the modulation of the periodic behavior, by the method of ``complex amplitude 
reconstruction'' described by \cite{1995AJ....109.1889F}.

In order to plot a realistic representation of the estimated spectral density when 
very closely spaced frequencies are involved, the discrete frequencies from a CLEANest 
analysis can be combined by convolving them with a ``clean beam'' in exactly the same 
manner as the treatment of the CLEAN spectrum by \cite{1987AJ.....93..968R}

For this analysis, we completed the CLEANest analysis for up to 12 statistically-signficant
frequencies. In most cases, fewer than six frequencies were significant.

\section{A New Modulated-Blazhko RR1: OGLE-BLG-RRLYR-6387}

OGLE-BLG-RRLYR-03825 is a ``modulated-Blazhko'' RR1 variable which shares a number of
characteristics with the only other known instance of its kind, the
galactic field RR1 star LS Her. LS Her's first overtone period is 0.2308078 d;
\cite{2008MNRAS.387..783W} found it to have a Blazhko period of 12.75 d which 
itself was modulated with a period of 109 d. The first overtone period of
OGLE-BLG-RRLYR-03825 is 0.2774114 d - based on the OGLE-III I-band photometry
alone - and our analysis found the star to have a Blazhko period  of 16.469 d
which is modulated with a period of 339.2 d. 

\begin{figure}[p]
\centering
\includegraphics[width=0.575\textwidth]{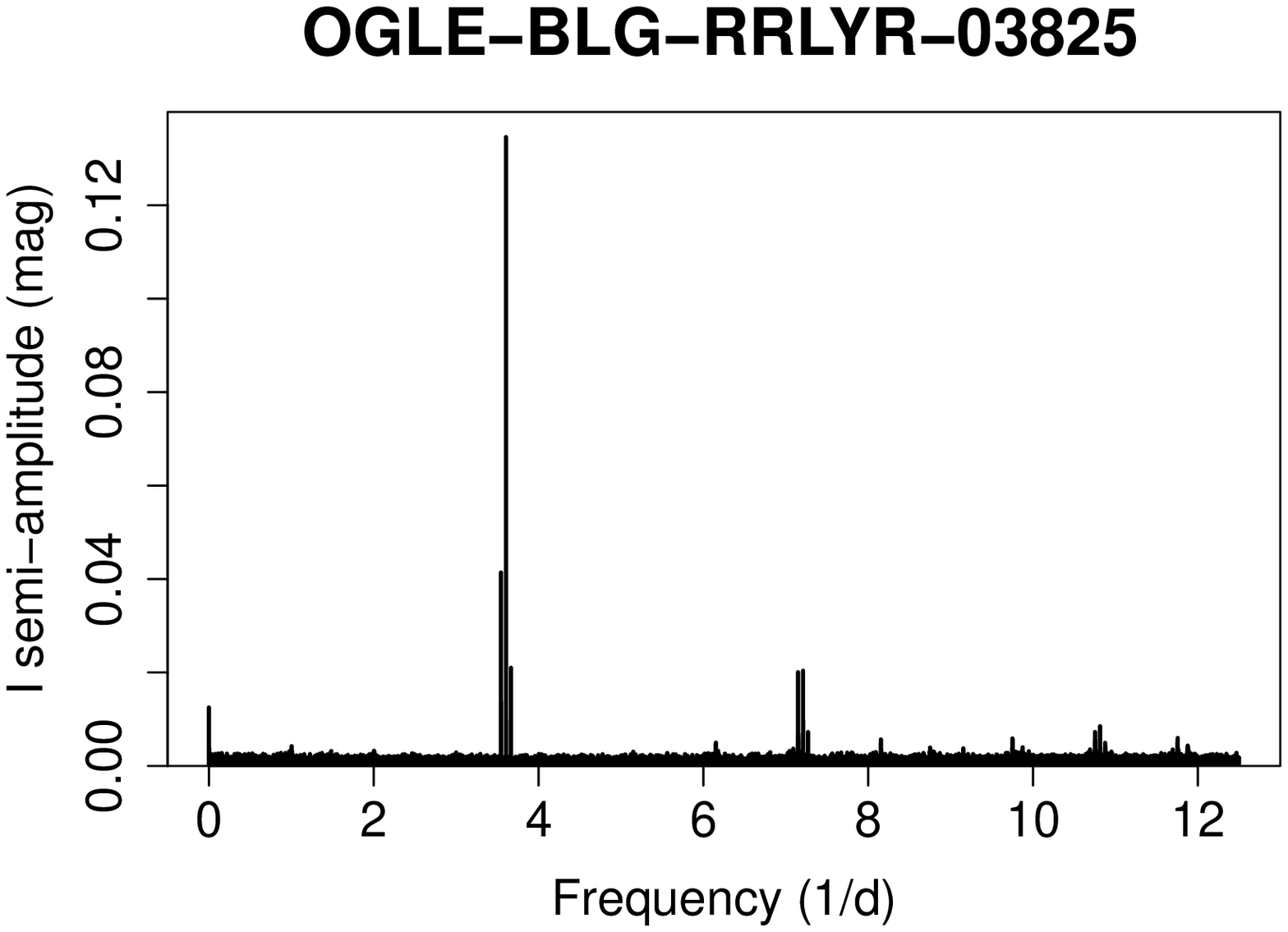}
\vskip0pt
\caption{The amplitude spectrum of OGLE-BLG-RRLYR-03825. Note that the fundamental (near
a frequency of 3.6 d$^{-1}$) and its first two harmonics have adjacent Blazhko peaks which are themselves modulated.}
\label{fig:ampfrq03825a}
\centering
\includegraphics[width=0.575\textwidth]{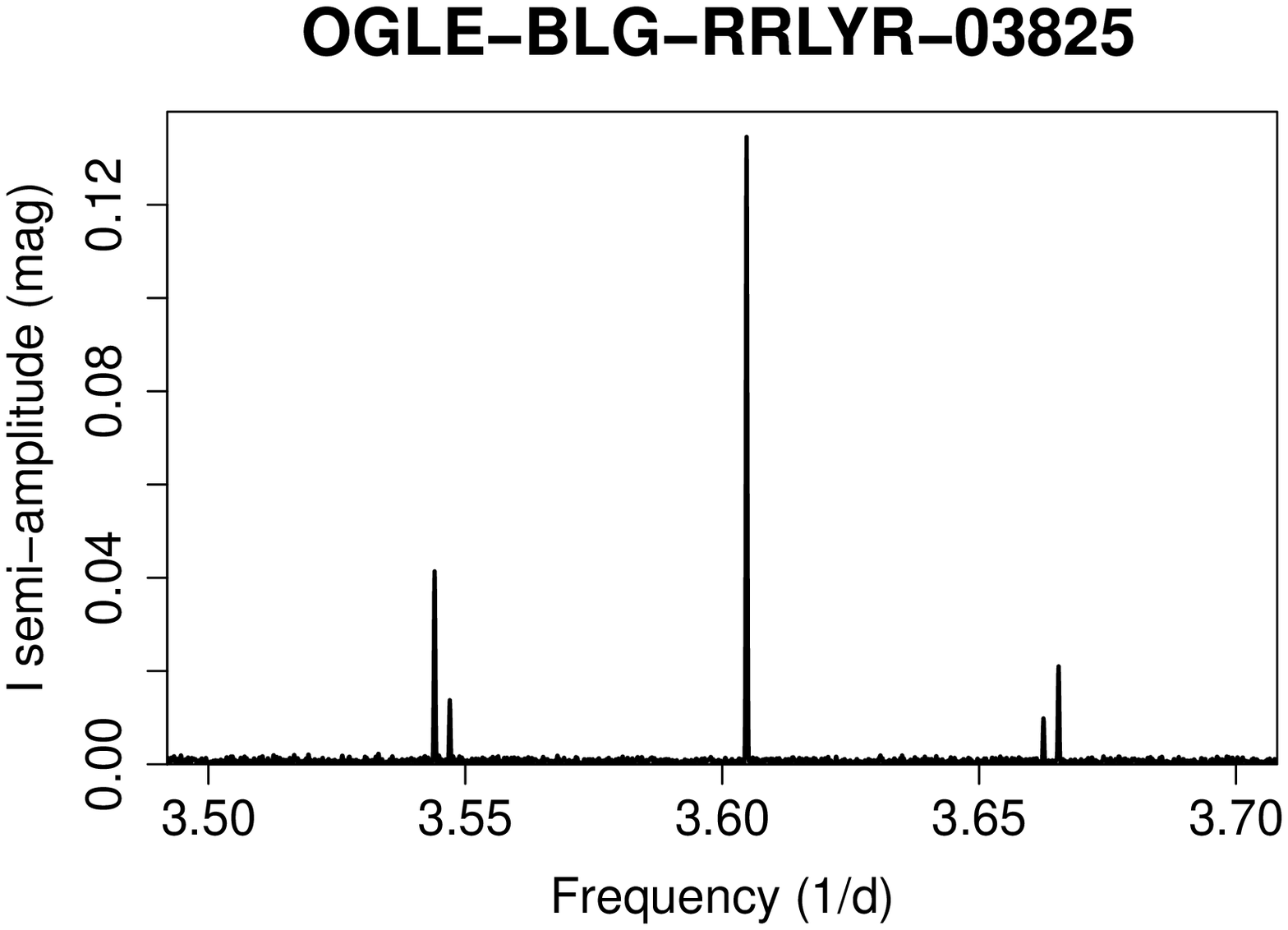}
\vskip0pt
\caption{A portion of the amplitude spectrum of OGLE-BLG-RRLYR-03825 in the
immediate vicinity of the fundamental mode peak near 3.6 d$^{-1}$, revealing apparently periodic
modulation of the Blazhko period itself.}
\label{fig:ampfrq03825b}
\end{figure}

The section of the amplitude spectrum of OGLE-BLG-RRLYR-03825 which is
wide enough to contain the fundmental frequency and its first two harmonics 
is shown in Figure~\ref{fig:ampfrq03825a}. In Figure~\ref{fig:ampfrq03825b},
a magnified section of the amplitude spectrum shows the modulation of the
Blazhko period itself.

\begin{figure}[p]
\centering
\includegraphics[width=0.575\textwidth]{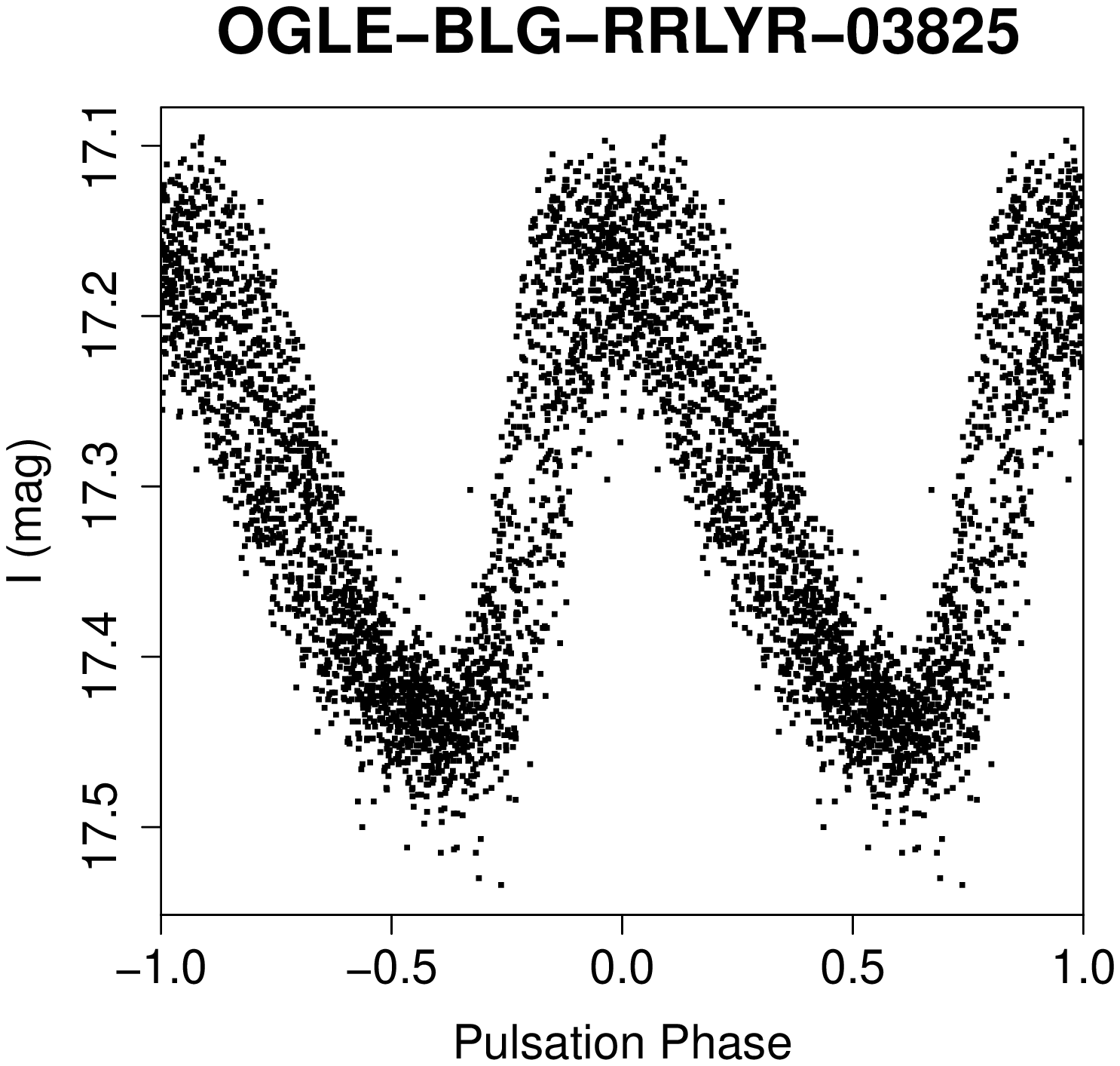}
\vskip0pt
\caption{The I-band lightcurve of OGLE-BLG-RRLYR-03825 phased with
the period 0.27741147 d.}
\label{fig:OGLE03825a}
\centering
\includegraphics[width=0.575\textwidth]{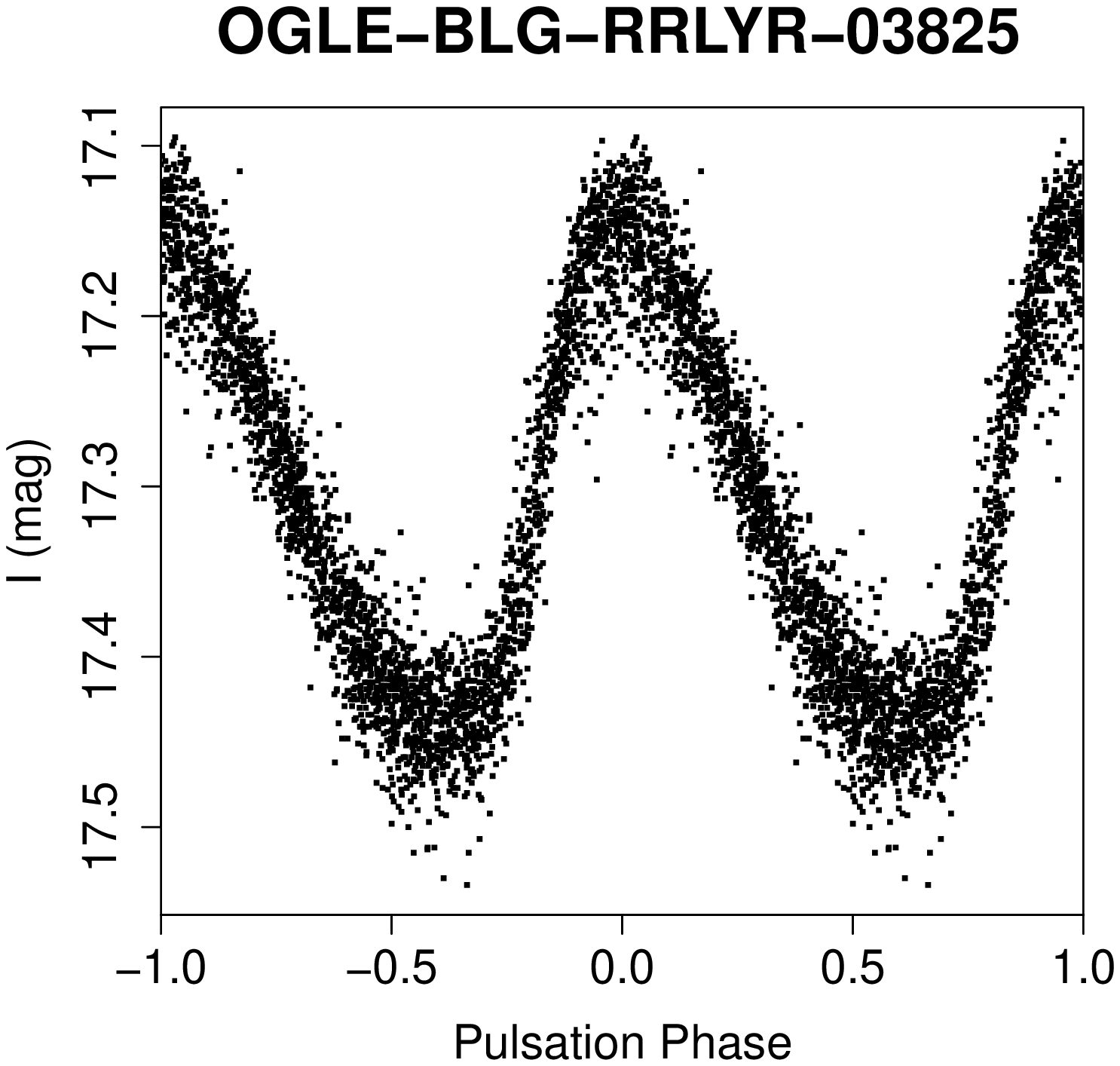}
\vskip0pt
\caption{The I-band lightcurve of OGLE-BLG-RRLYR-03825 with the time-varying
phase removed as determined by clusters of nearby frequencies
in a CLEANest model.}
\label{fig:OGLE03825b}
\end{figure}

\begin{figure}[p]
\centering
\includegraphics[width=0.575\textwidth]{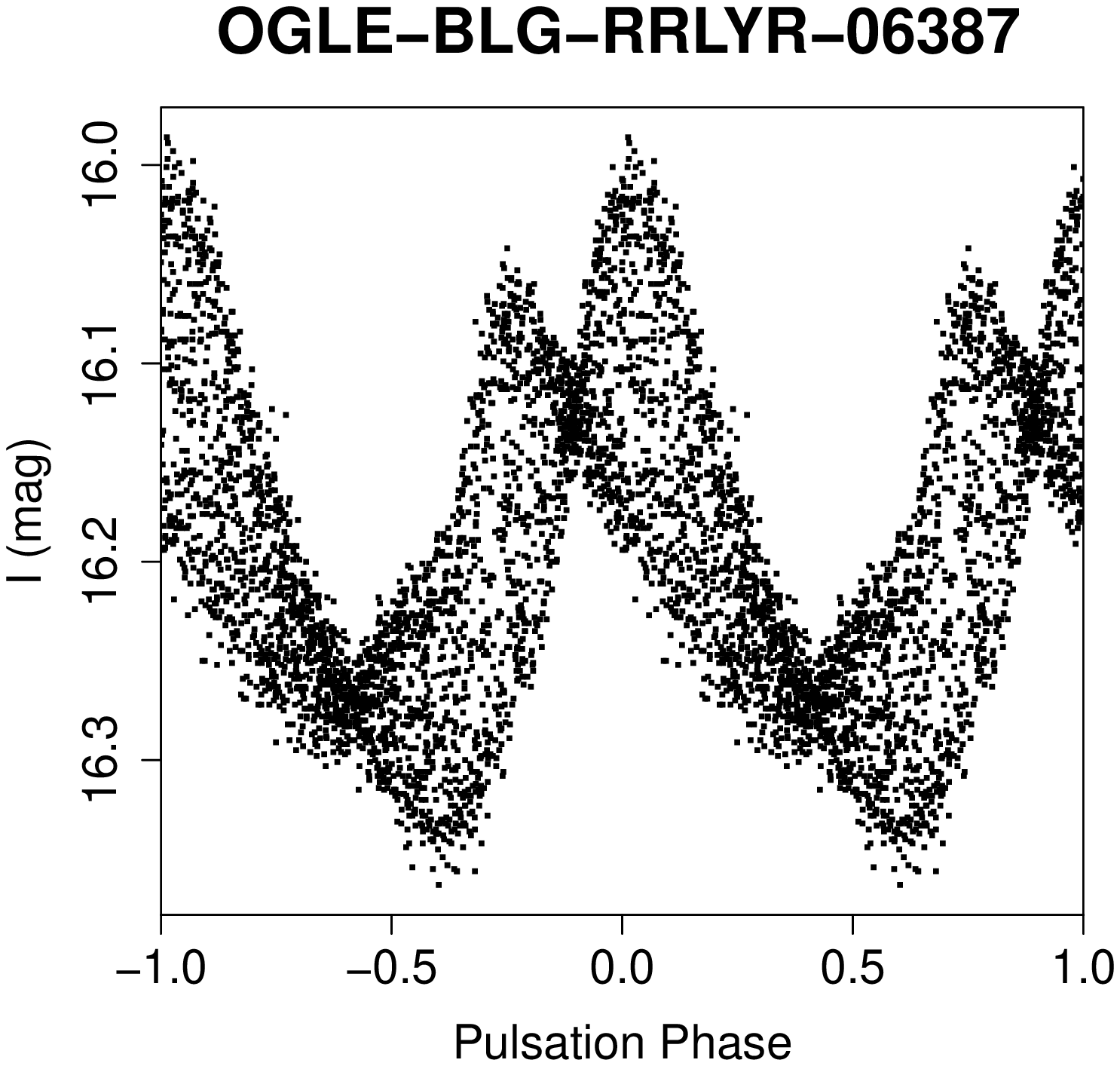}
\vskip0pt
\caption{The I-band lightcurve of OGLE-BLG-RRLYR-06387 phased with
the period 0.2717559 d.}
\label{fig:OGLE06387a}
\centering
\includegraphics[width=0.575\textwidth]{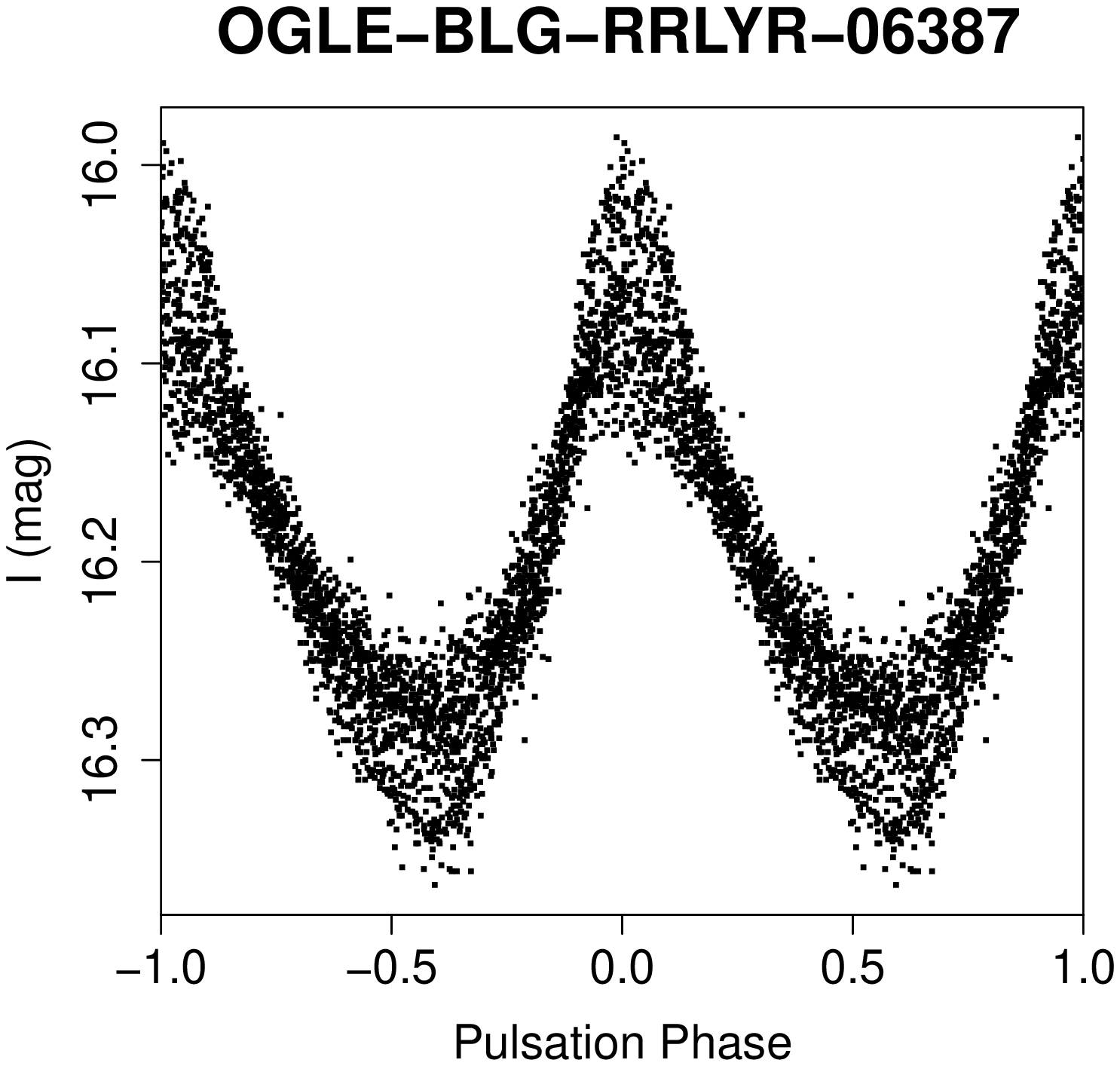}
\vskip0pt
\caption{The I-band lightcurve of OGLE-BLG-RRLYR-06387 with the time-varying
phase removed as determined by clusters of nearby frequencies
in a CLEANest model.}
\label{fig:OGLE06387b}
\end{figure}

\begin{figure}[p]
\centering
\includegraphics[width=0.575\textwidth]{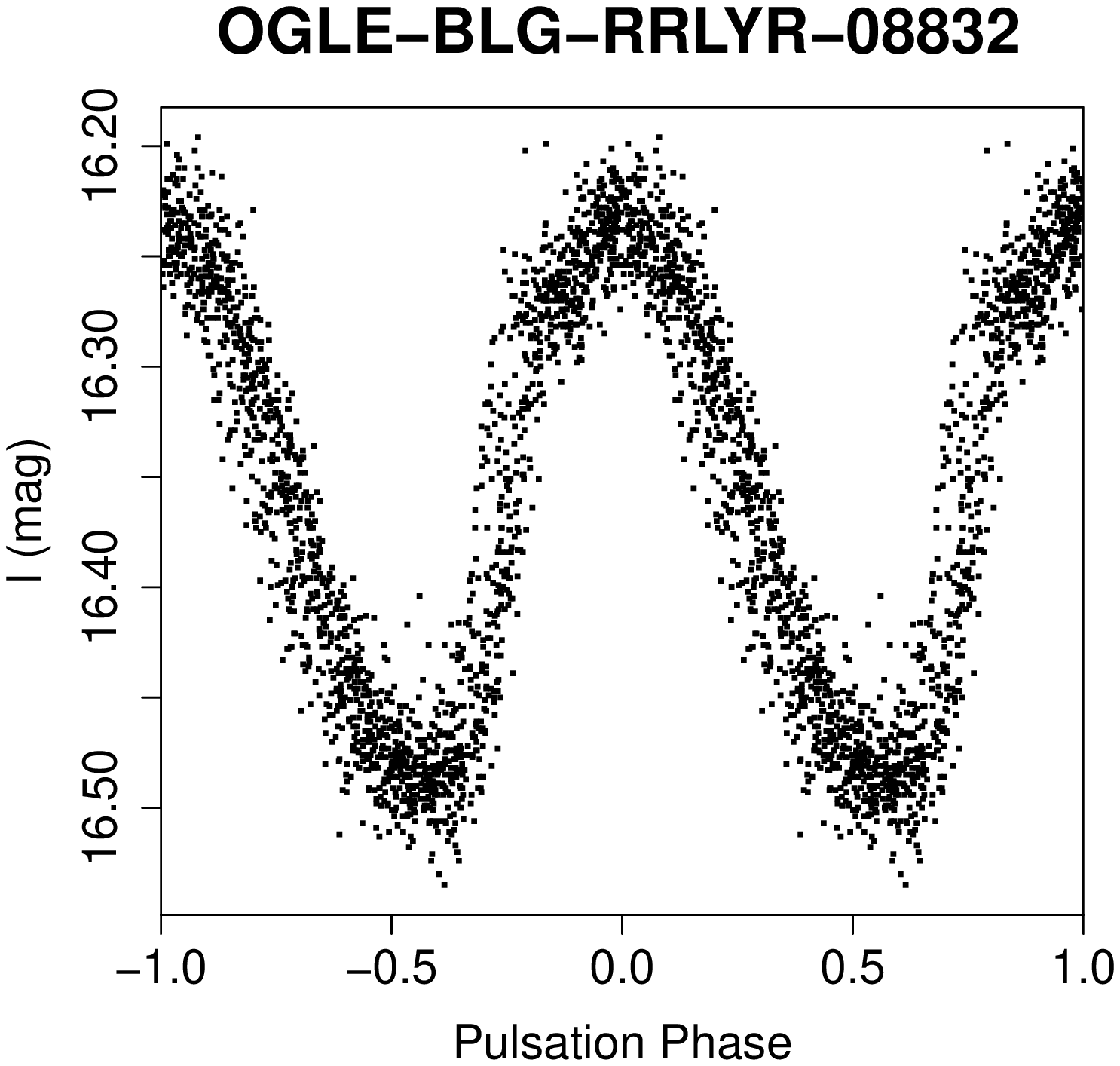}
\vskip0pt
\caption{The I-band lightcurve of OGLE-BLG-RRLYR-08832 phased with
the period 0.3036124 d.}
\label{fig:OGLE08832a}
\centering
\includegraphics[width=0.575\textwidth]{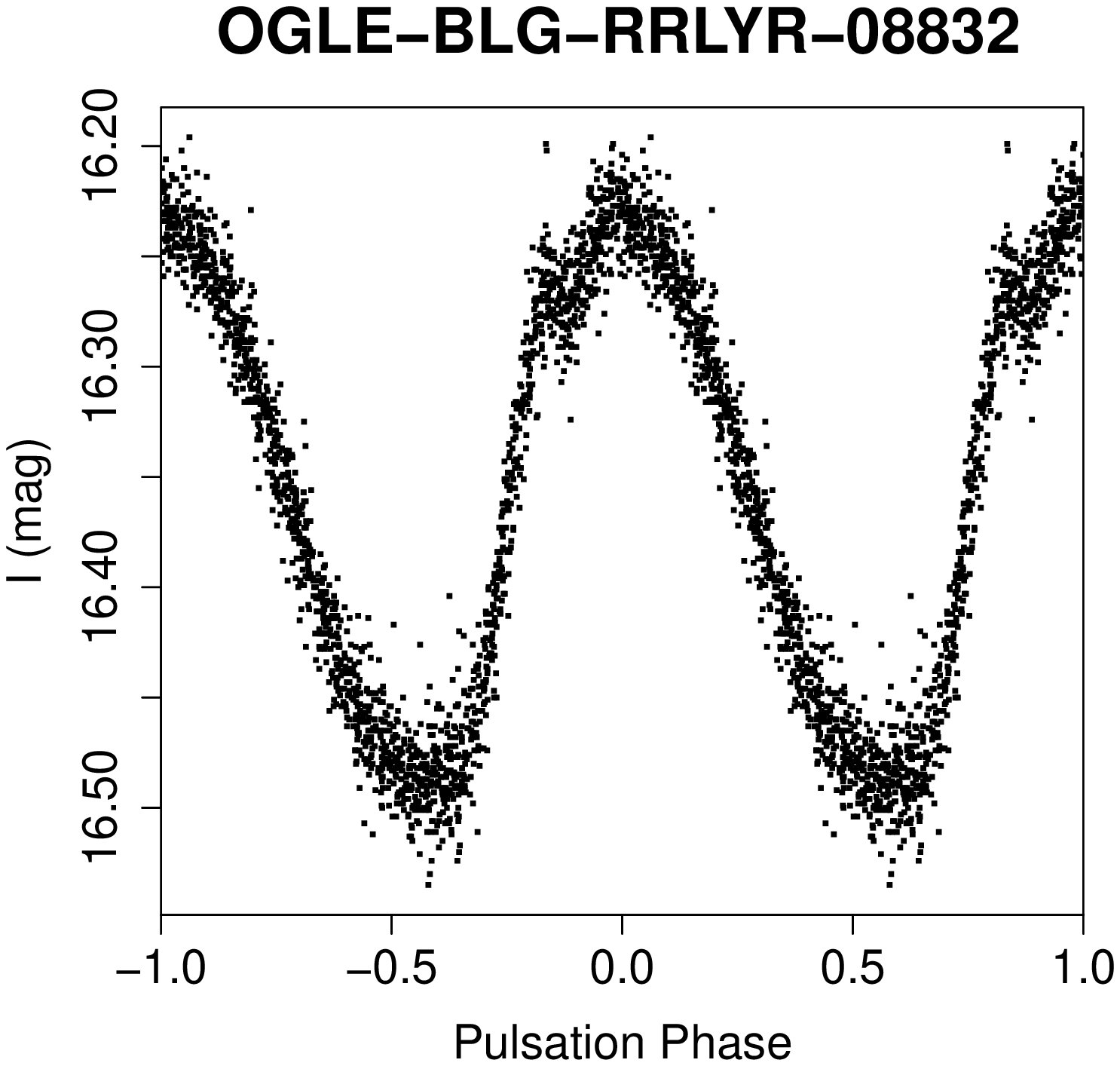}
\vskip0pt
\caption{The I-band lightcurve of OGLE-BLG-RRLYR-08832 with the time-varying
phase removed as determined by clusters of nearby frequencies
in a CLEANest model.}
\label{fig:OGLE08832b}
\end{figure}

\begin{figure}[p]
\centering
\includegraphics[width=0.575\textwidth]{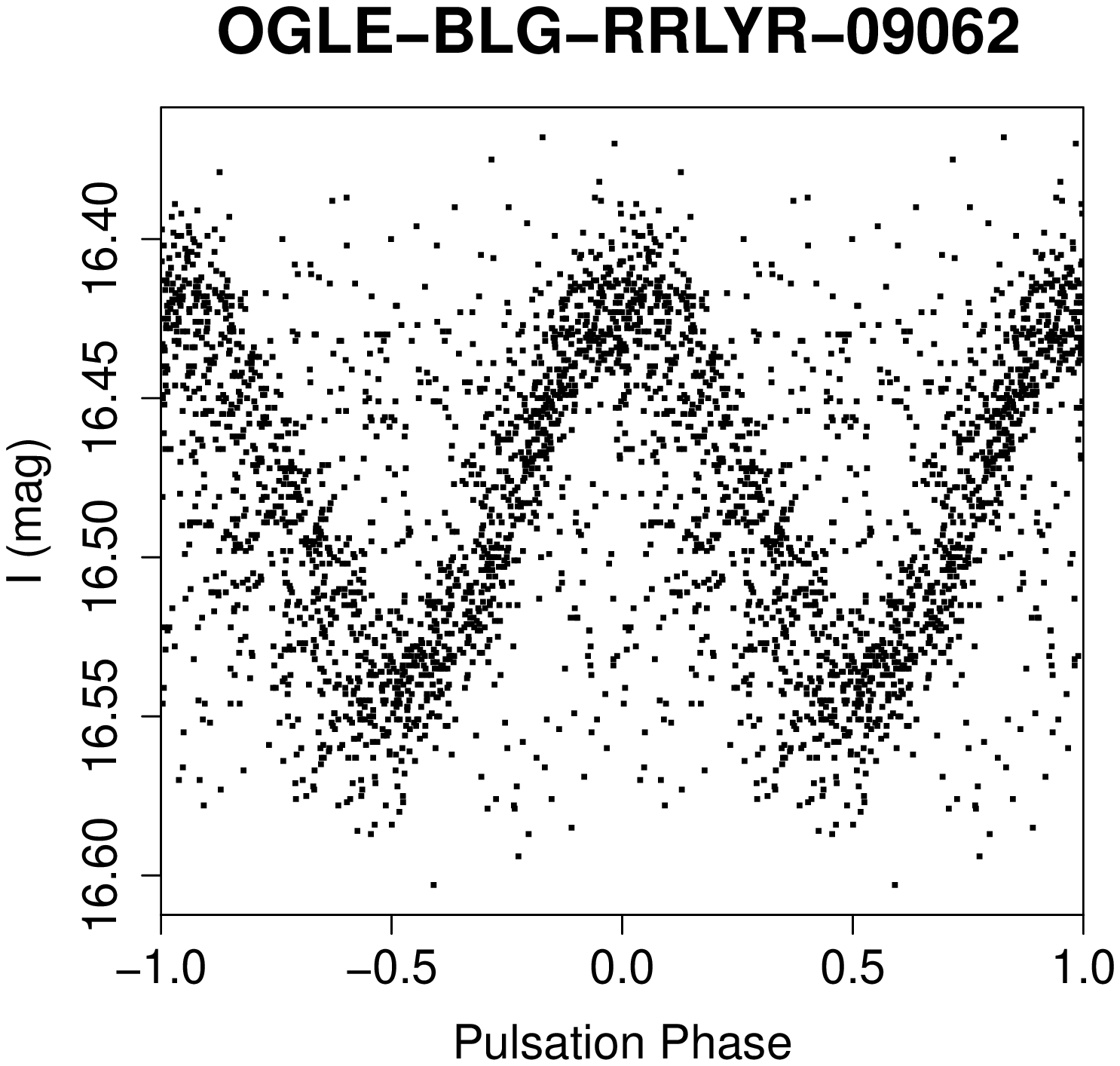}
\vskip0pt
\caption{The I-band lightcurve of OGLE-BLG-RRLYR-09062 phased with
the period 0.25949368 d.}
\label{fig:OGLE09062a}
\centering
\includegraphics[width=0.575\textwidth]{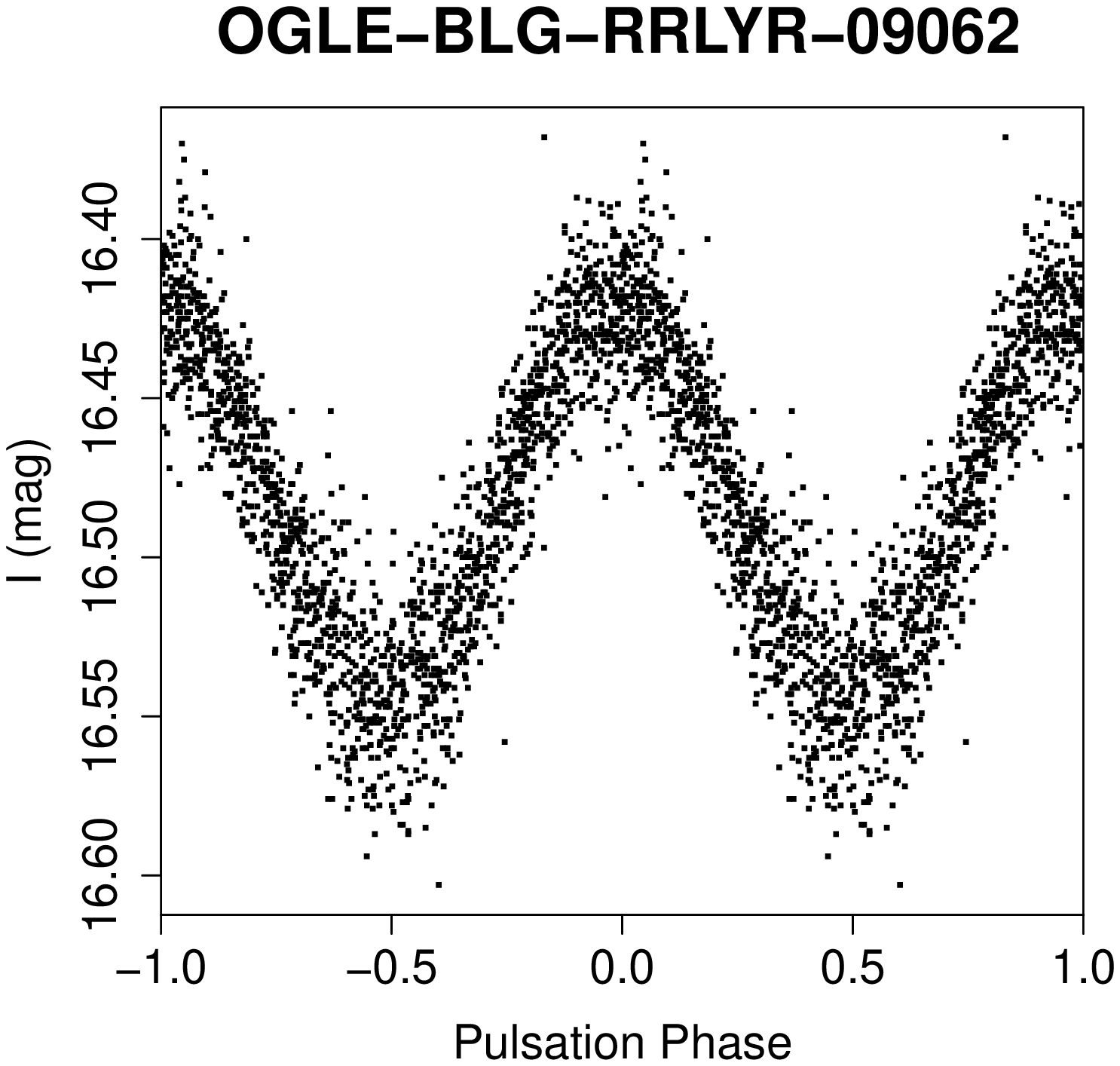}
\vskip0pt
\caption{The I-band lightcurve of OGLE-BLG-RRLYR-09062 with the time-varying
phase removed as determined by clusters of nearby frequencies
in a CLEANest model.}
\label{fig:OGLE09062b}
\end{figure}

\begin{figure}[p]
\centering
\includegraphics[width=0.575\textwidth]{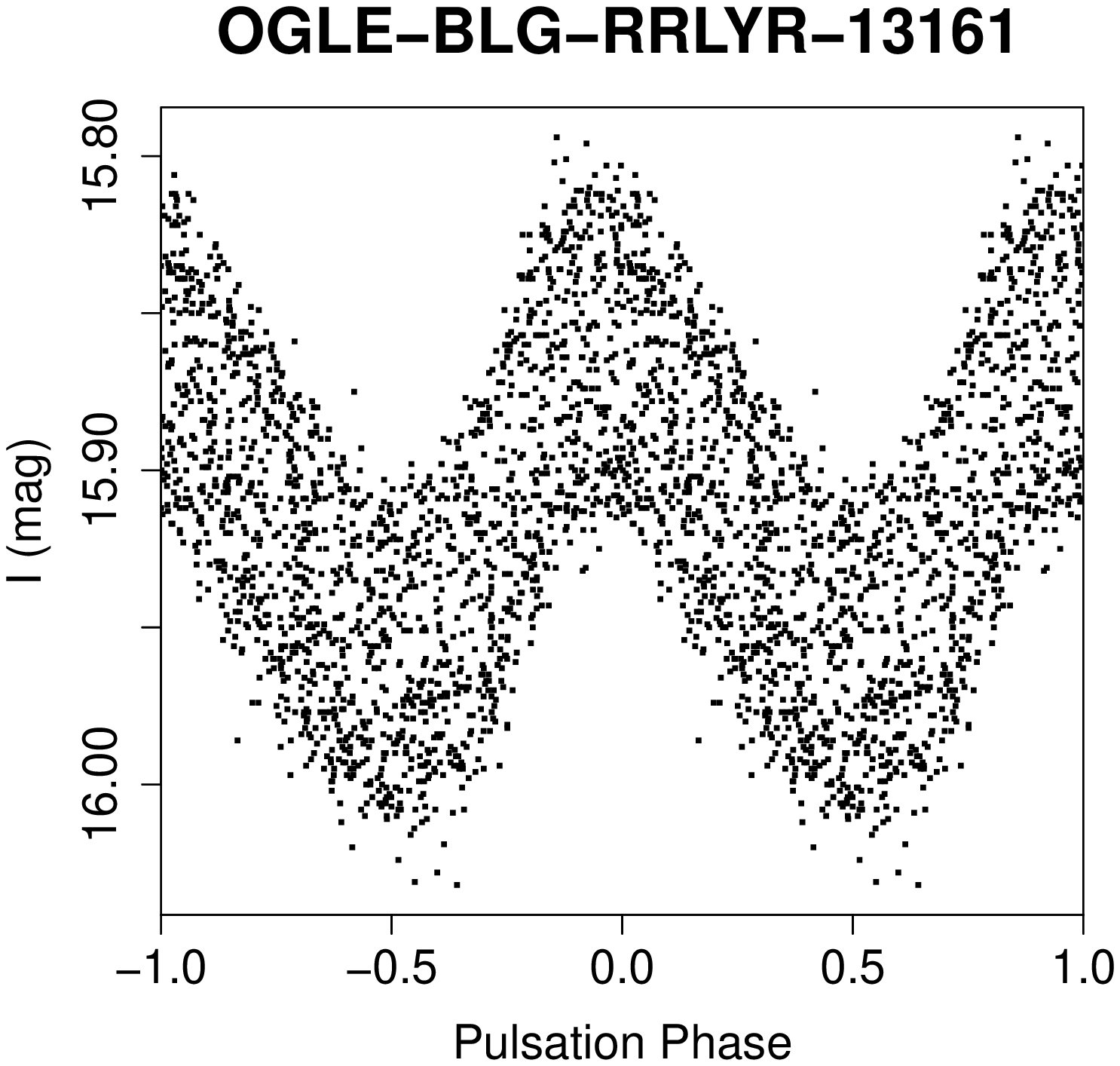}
\vskip0pt
\caption{The I-band lightcurve of OGLE-BLG-RRLYR-13161 phased with
the period 0.25375814 d.}
\label{fig:OGLE13161a}
\centering
\includegraphics[width=0.575\textwidth]{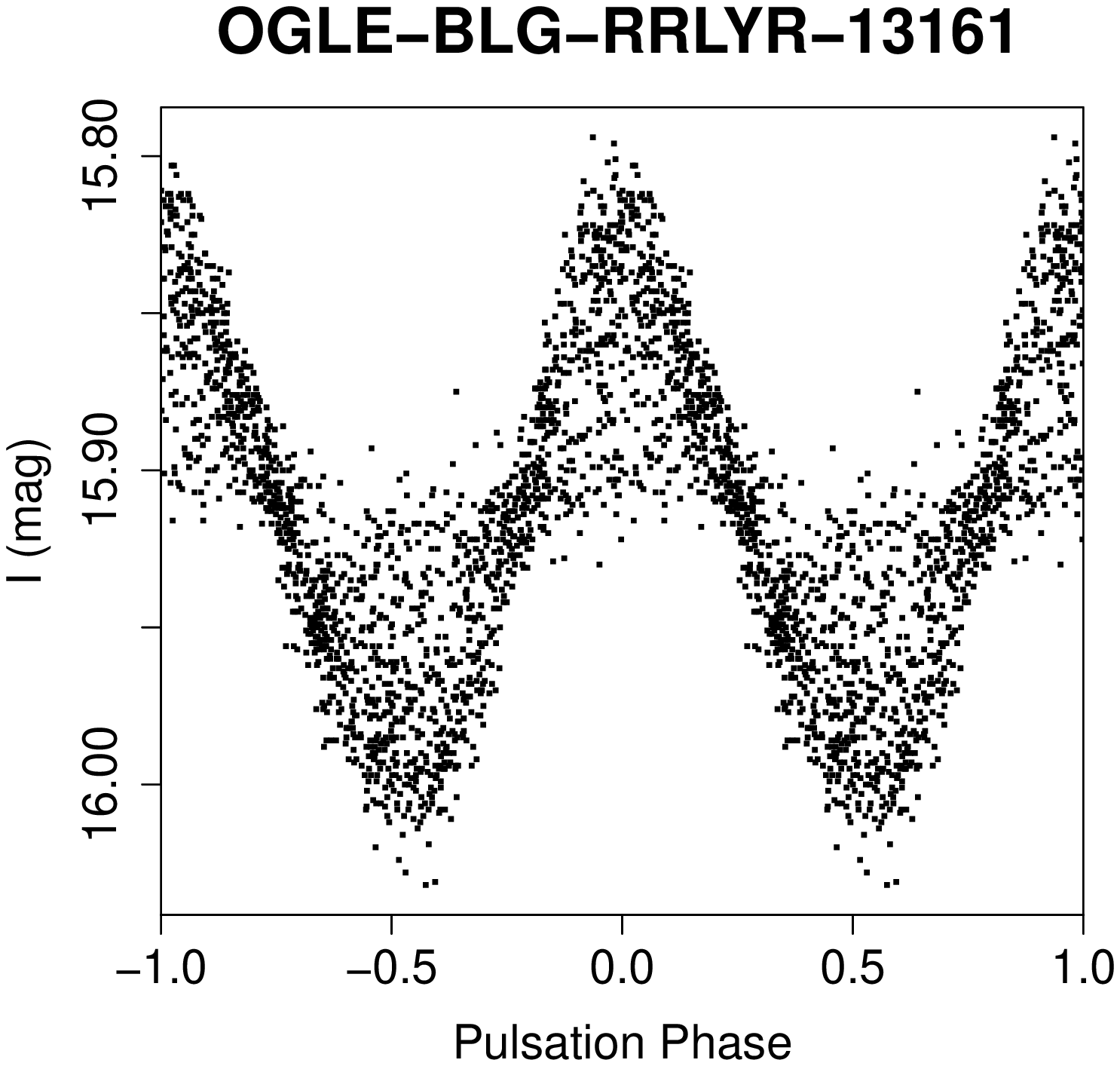}
\vskip0pt
\caption{The I-band lightcurve of OGLE-BLG-RRLYR-13161 with the time-varying
phase removed as determined by clusters of nearby frequencies
in a CLEANest model.}
\label{fig:OGLE13161b}
\end{figure}

Figures~\ref{fig:OGLE03825a} and \ref{fig:OGLE03825b} display the lightcurve
phased with constant period and with the time-varying phase determined by
our analysis removed, respectively. Removal of the time-varying phase
preserves the underlying lightcurve shape for Fourier coefficient analysis.
The improvement in the definition of the lightcurve shape by removing 
the time-varying phase is demonstrated for several other modulated lightcurves
in Figures~\ref{fig:OGLE06387a}, \ref{fig:OGLE06387b}, \ref{fig:OGLE08832a},
\ref{fig:OGLE08832b}, \ref{fig:OGLE09062a}, \ref{fig:OGLE09062b}, \ref{fig:OGLE13161a},
and \ref{fig:OGLE13161b}, respectively.

\cite{2008MNRAS.387..783W} noted it is possible that other galactic field RR1
stars have modulated-Blazhko lightcurves but that the duration and quality
of most existing time-series would not allow them to be easily detected.
The OGLE-III bulge data is more uniform in this regard and yet this
object was one of only a few modulated-Blazhko candidates, suggesting the occurrence
of stars with detectable power (from ground-based surveys) in Blazhko-modulated 
variation is low - less than 2\%.

\section{Conclusions and Future Work}
The principal findings of our analysis to date are:
\begin{itemize}
\item Lightcurve amplitude and phase modulation among RR1 stars is
function of period and we find that shorter-period stars have significantly
higher rates of such activity.
\item A ``modulated-Blazhko'' star has been found -
OGLE-BLG-RRLYR-03825. Like the galactic field RR1 star LS Her, it is a short-period
(0.2774114 d) RR1 and the Blazhko period is short (16.469 d). The short-period 
Blazhko modulation itself is modulated with a 339.2 d period.
\item We have seen no clear evidence of binarity to date. This is 
possibly due to the higher rates of non-periodic lightcurve
modulation in RR1 stars swamping any existing signal due to binarity.
\item Our unbiased selection of significant frequencies revealed no clear period-doubling
candidates in this sample. We note that the semi-amplitudes of the period-doubling
peaks detected by \cite{2012arXiv1208.4251M} are similar to the detection threshold
of our ground-based time-series, so this finding sets an upper limit for their
typical power, if such peaks indeed exist in most RR1 stars.
\end{itemize}

We are continuing the analysis of the frequency list derived from the well-observed
sample. Future opportunities for greater understanding from this rich dataset
include the search for peculiar period ratios, such as those found in the only four known
RR1 stars in the Kepler field by \cite{2012arXiv1208.4251M}, and searching for correlations between
lightcurve behavior and metallicity using relations similar to those
derived by \cite{2007MNRAS.374.1421M}, but adapted for I-band lightcurves.

\section*{Acknowledgments}
The research of DW is supported, in part, by a Discovery Grant from NSERC (Natural Sciences
and Engineering Research Council of Canada).

\end{document}